\documentclass[journal=nalefd,manuscript=letter]{achemso}
\usepackage{bm}
\usepackage{multirow}
\usepackage{rotating}
\usepackage{mciteplus}
\usepackage{footmisc}

\author{Liujiang Zhou}
\affiliation{Bremen Center for Computational Materials Science, University of Bremen, Am Falturm 1, 28359 Bremen, Germany}
\alsoaffiliation{These authors contributed equally to this work}
\email{liujiang86@gmail.com}
\author{Jin Zhang}
\affiliation{Beijing National Laboratory for Condensed Matter Physics, and Institute of Physics, Chinese Academy of Sciences, Beijing, 100190, P. R. China}
\alsoaffiliation{These authors contributed equally to this work}
\author{Zhiwen Zhuo}
\affiliation{Department of Materials Science and Engineering, University of Science and Technology of China, Hefei, Anhui 230026, China}
\author{Liangzhi Kou}
\affiliation{School of Chemistry, Physics and Mechanical Engineering Faculty, Queensland University of Technology, Garden Point Campus, QLD 4001, Brisbane, Australia}
\author{Wei Ma}
\affiliation{Beijing National Laboratory for Condensed Matter Physics, and Institute of Physics, Chinese Academy of Sciences, Beijing, 100190, P. R. China}
\author{Bin Shao}
\affiliation{Bremen Center for Computational Materials Science, University of Bremen, Am Falturm 1, 28359 Bremen, Germany}
\author{Aijun Du}
\affiliation{School of Chemistry, Physics and Mechanical Engineering Faculty, Queensland University of Technology, Garden Point Campus, QLD 4001, Brisbane, Australia}
\author{Sheng Meng}
\affiliation{Beijing National Laboratory for Condensed Matter Physics, and Institute of Physics, Chinese Academy of Sciences, Beijing, 100190, P. R. China}
\alsoaffiliation{Collaborative Innovation Center of Quantum Matter, Beijing 100190, China}
\email{smeng@iphy.ac.cn}
\author{Thomas Frauenheim}
\affiliation{Bremen Center for Computational Materials Science, University of Bremen, Am Falturm 1, 28359 Bremen, Germany}

\title[An \textsf{achemso} demo]
 {Phosphorene and Doped Monolayers Interfaced TiO$_2$ with Type-II Band Alignments: Novel Excitonic Solar Cells}
\abbreviations{phosphorene, van der Waals heterostructure, doping effect, exciton binding energy, excitonic solar cell, charge separation, electron dynamics}
\keywords{phosphorene, van der Waals heterostructure, doping effect, exciton binding energy, excitonic solar cell, charge separation, electron dynamics}

\begin{document}
\begin{abstract}
Phosphorene, a new elemental two dimensional (2D) material recently isolated by mechanical exfoliation, holds the feature of a direct band gap of around 2.0 eV, overcoming graphene's weaknesses (zero band gap) to realize the potential application in optoelectronic devices. Constructing van der Waals  heterostructures is an efficient approach to modulate the band structure, to advance the charge separation efficiency, and  thus to optimize the optoelectronic properties. Here, we theoretically investigated three type-II heterostructures based on perfect phosphorene and its doped monolayers interfaced with TiO$_2$(110) surface. Doping in phosphorene  has a tunability on built-in potential, charge transfer, light absorbance, as well as electron dynamics, which helps to optimize the light absorption efficiency. Three excitonic solar cells (XSCs) based on the phosphorene$-$TiO$_2$ heterojunctions have been proposed, which exhibit high power conversion efficiencies dozens of times higher than conventional solar cells, comparable to MoS$_2$/WS$_2$ XSC. The nonadiabatic molecular dynamics within the time-dependent density functional theory framework shows ultrafast electron transfer time of 6.1$-$10.8 fs, and slow electron$-$hole recombination of 0.58$-$1.08 ps, yielding $>98\%$ quantum efficiency for charge separation, further guaranteeing the practical power conversion efficiencies in XSC.

\end{abstract}

Although graphene shows intriguing electronic and mechanical properties and is highly desired in the field of next generation of faster and smaller electronic devices,\cite{tombros_electronic_2007,ohta_controlling_2006} the feature of zero band gap makes it unsuitable for the controlled and reliable transistor operation and consequently limits its widespread applications in optoelectronic devices, such as light-emitting diodes, field effect transistors, and solar cells.\cite{zhou_sic2_2013} It is therefore highly desired to open an energy gap in graphene. Subsequently, a whole new class of 2D materials has been studied and prepared in experiment, such as the transition metal dichalchogenides,\cite{xu_graphene-like_2013,splendiani_emerging_2010, mak_tightly_2013,britnell_strong_2013} transition-metal carbides (MXenes),\cite{wang_atomic-scale_2015,anasori_two-dimensional_2015} monolayer black phosphorus (BP) (termed phosphorene),\cite{qiao_high-mobility_2014} etc.  Among them, phosphorene, a single layer of BP arranged in a hexagonal puckered lattice,  isolated from the bulk black phosphorus structure, has demonstrated extraordinary  properties,  including its highly anisotropic effective masses, high electron mobility ($>$ 1000 cm$^2$/V$\cdot$s), a direct band gap of  $\sim$2.0 eV.\cite{qiao_high-mobility_2014, li_black_2014} The band structure and related optoelectronic properties can be modulated through a variety of ways, such as doping,\cite{low_tunable_2014, kim_observation_2015,hashmi_transition_2015} point and line defects,\cite{liu_two-dimensional_2014} layer-by-layer\cite{tran_layer-controlled_2014} or hetero-structures mapping,\cite{padilha_van_2015,yuan_photoluminescence_2015} alloys forming, chemical adsorption,\cite{ziletti_oxygen_2015} strain,\cite{fei_strain-engineering_2014} and  external electric field,\cite{koenig_electric_2014} broadening its widespreading applications. These features  are complementary to the gapless graphene, suggestive of a great potential in optoelectronic applications, including sensors, modulators, solar cells, and light-emitting diodes (LEDs).\cite{buscema_photovoltaic_2014,lam_performance_2014,li_black_2014}

Van der Waals (vdW) heterostructure formed by stacking 2D atomic monolayers, is one of the best approaches to protect the active layers against environmental contamination without affecting their electrical performance, or to modulate the band offsets at the interfaces so as to provide a highly effective
means for the manipulation of charge carriers.\cite{buscema_photocurrent_2015,wang_van_2015} Vdw hetero-structures  can be widely used  to enhance the electron-holes separation when used as  an excitonic solar cells (XSC). In the XSC, power conversion efficiency (PCE) depends critically on the interface band alignment between donor and acceptor materials.  The type-II band alignment between interfaces  is  a prerequisite to achieve the efficient electron-hole pairs (excitons) separation, which has been achieved in various 2D layered transition metal dichalchogenides-,\cite{bernardi_extraordinary_2013} carbon- or silicon-based  heterostructures,\cite{zhou_sic2_2013} etc. Although such phosphorene-based hetero-structures are widely reported in experiment and theory, type-II band alignment heterostructures are extremely scarce, only being BP/MoS$_2$\cite{huang_electric-field_2015, yuan_photoluminescence_2015,chen_gate_2015}, BP/GaAs\cite{gehring_thin-layer_2015} as well as black$-$red phosphorus heterostructures.\cite{shen_blackred_2015}. Therefore, desirable type-II phosphorene-based heterostructures preferably with a large built-in potential for driving electrons and holes deserve to be explored in experiment and theory.

TiO$_2$ possesses normally a wide band gap and excellent electrical conductance, and has been widely used as a promising photocatalysis and photovoltaic  materials,\cite{ni_review_2007,zhang_importance_2008,gratzel_photoelectrochemical_2001} due to the fact that the generated photoexcited electrons (PEs) can either be readily channeled to create electricity directly in solar cells or be used to drive water splitting for hydrogen production. However, the large energy gap of pristine TiO$_2$ (3.0 eV for rutile and 3.2 eV for anatase) limits its actual efficiency to generate  photoexcited electrons and the subsequential dissociation of excitons. With these above factors in mind, interfacing wide-band-gap TiO$_2$  with phosphorene may offer a straightforward approach to overcome these shortcomings in light absorption, so as to  increase the light absorption efficiency via harvesting solar spectrum in a wider frequency range, and also to enhance the exciton separation efficiency due to intrinsic conduction band offsets  in the region of interfaces.  Layered BP interfaced with TiO$_2$ nanoparticles with a type-II band alignment shows novel photocatalytic performances  over  traditional graphene-TiO$_2$ hybrid.\cite{uk_lee_stable_2015, wang_electronics_2012} However, studies of phosphorene-TiO$_2$ hybrids, especially interfaced with TiO$_2$ crystals, on photovoltaic application are still unexplored. Due to the low level of valence band maximum (VBM) of TiO$_2$, phosphorene interfaced with TiO$_2$ (1L-BP@TiO$_2$) may form an efficient XSC where phosphorene is used as electron donor and TiO$_2$ as electron acceptor.

By utilizing the large-scale density functional theory and GW + Bethe-Salpeter equation (BSE) calculations , we first systemically investigate three heterostructures consisting of phosphorene or Al-doped (hole-doped) or Cl-doped (electron-doped) monolayers interfaced with TiO$_2$(110) surface (Figure 1a). Our calculations reveal that these heterojunctions show type-II band alignments, and  large built-in potential for carrier separation, which can efficiently mediate the direct electron excitation from phosphorene into the titania conduction band (CB) under visible light irradiation, indicating a dramatic enhanced photoactivity  in low-energy region of 1.0$-$2.5 eV. Doping in phosphorene has a well tunability on the excitonic binding energy, optical band gap, and light absorbance, which helps to optimize the light absorption in active layers. We have proposed three XSCs based on perfect or doped monolayers interfaced with TiO$_2$(110) surface, which show tunable total power conversion efficiency of about 1.5\% and ultrahigh power densities of $\sim$12.2$-$17.1 MW/L, dozens of times higher than conventional GaAs (290 kW/L) and Si (5.9 kW/L) solar cells, and comparable with MoS$_2$/WS$_2$ XSC with 1nm thickness. Nonadiabatic molecular dynamics simulation provides important insights into the charge-separation and electron$-$hole recombination processes, showing an ultrafast electron transfer time of 6.1$-$10.8 fs, and slow electron$-$hole recombination time of 0.58$-$1.08 ps, further guaranteeing the practical solar power conversion efficiencies. The structure, electronic and optical properties and  dynamics for charge-separation and electron$-$hole recombination processes are discussed in detail.


Our design of heterostructure places a single-layer phosphorene on rutile TiO$_2$(110) surface (1L-BP/TiO$_2$(110)). Figure 1b and c show the optimized 1L-BP/TiO$_2$(110) heterostructure at the atomical level. The principal idea of doping Al (hole-doping) or Cl (electron-doping) atoms in phosphorene layers interfaced with TiO$_2$(110) surface (denoted as Al@1L-BP/TiO$_2$(110) and Cl@1L-BP/TiO$_2$(110), respectively) may increase or reduce the interaction within interfaces and therefore to modulate the electron transfer. Doping enhanced interaction between interfaces can be verified by the shortened P$-$O bond distances of 2.67 and 2.71 $\mathrm{\AA}$  in Al- and Cl-doped phosphorene, 0.14 and 0.1 $\mathrm{\AA}$  shorter than perfect phosphorene (2.81 $\mathrm{\AA}$) respectively. The equilibrium distance between the phosphorene and the top of the TiO$_2$(110) surface are calculated to be 2.75 \AA. As for Al@1L-BP/TiO$_2$(110) and Cl@1L-BP/TiO$_2$(110), the interlayer distances have a decrease by 0.4 and 0.2 \AA, respectively,  indicating doping would strengthen the interlayer interaction compared to 1L-BP/TiO$_2$(110).

The interlayer coupling interaction can also be assessed by the assessment of the extent of charge transfer. In principle, the more charge transfers from phosphorene layers to TiO$_2$, the more stronger interlayer coupling is. To clarify the charge transfer and separation processes, 3D charge density difference plots are calculated by subtracting the electronic charge of a hybrid phosphorene/TiO$_2$ nanocomposite from the separate phosphorene layers and TiO$_2$(110) surface.  Figure 1d$-$f show there are obvious charge transfers from phosphorene layers to the TiO$_2$(110) surface in the ground electronic state in the three interfaces, implying efficient hole accumulation in the titania-supported phosphorene layers. Charge redistribution mostly occurs at the phosphorene/TiO$_2$(110) interface region, while there is almost no charge transfer on the TiO$_2$ substrate farther from the interface. Upon formation of an interface, the electrons are able to excite from the ground states of phosphorene layers to their excited states and then are transferred to the TiO$_2$(110) conduction band (CB), leading to efficient electron-hole pair separation.  More importantly,  the amount of charge transfer in Al@1L-BP/TiO$_2$ interfaces is greater than that in 1L-BP/TiO$_2$ and Cl@1L-BP/TiO$_2$, due to the much stronger donor-acceptor interaction.  In an effort to quantitatively estimate the amount of charge transfer between the phosphorene layers and the TiO$_2$, we have performed the Bader charge analysis.\cite{henkelman_fast_2006} The averaged Bader charge states show that perfect phosphorene, Al@1L-BP and Cl@1L-BP lose about  0.151, 0.162 and 0.105 electron, respectively, while TiO$_2$(110) surfaces gains the corresponding electron, which is consistent with the above difference charge density analysis shown in Figure 1c$-$d.  At the same time, we have also taken note that the total amount of charge transfer is much less than 1.0 electron from phosphorene layers to TiO$_2$, indicating the interfaces is weakly coupled  via the vdW interaction.

The large amount of electron transfer under light irradiation between interfaces can be attributed to the large built-in potential. A small built-in potential can not effectively drive the electron-hole pair separation from the interface region. Therefore, a large built-in potential is a prerequisite to drive charge carriers, resulting in effective separation, and finally to achieve high power conversion efficiency (PCE) in a XSC. To get the intrinsic built-in potentials and photoactivity for three heterostructures,  projected density of states (PDOSs) of three interfaces  are calculated. As shown in Figure 2a, c and Figure S1, the electronic states of  phosphorene or doped monolayers and TiO$_2$(110) do not hybridize together near the Fermi energy ($E_F$), almost remaining the same because donor-acceptor coupling is not strong enough in the absence of covalent bonding.  We take 1L-BP/TiO$_2$(110) as an example, large phosphorene states are localized within the TiO$_2$(110) gap, indicating the dominant electron transition is from O$-2$p states at the valence band (VB) to Ti-3d states at the CB under ultraviolet (UV) irradiation. More importantly, the electron can be excited from the first or second van Hove singularity of phosphorene to its CB under visible light irradiation, and then, this photogenerated electron (PE) is injected into the CB of TiO$_2$. The three interfaces with effective band gaps of 0.26, 0.35 and 0.36 eV, respectively, show type-II band alignment (Figure 2b).  Al@1L-BP/TiO$_2$ has a built-in potential (CBM offsets, denoted as $\Delta$$\emph{E}_C$) of about 0.4 eV, larger than 1L-BP/TiO$_2$(110) ($\sim$0.25 eV) and Cl@1L-BP/TiO$_2$(110) ($\sim$0.3 eV). The large CBM  (0.2$-$0.4 eV) and 1.8 eV VBM offsets ($\Delta$$\emph{E}_V$) can drive electrons to migration from phosphorene CBM to TiO$_2$ CBM, and hole transfer between their VBMs, respectively, indicating the highly efficient charge separation. These features, including type-II band alignments, large built-in potentials, enable all of three heterostructures to hold the potential capabilities of forming highly effective excitonic solar cells (XSCs).

The charge-transfer complex formed at interfaces is also expected to mediate photocatalytic activities under visible light. To confirm this effect, the independent particle absorbance spectrums of 1L-BP/TiO$_2$(110) and TiO$_2$(110) substrate are calculated with the random phase approximation (DFT-RPA) due to the large supercell, as shown in Figure 2d. The optical absorption for pure TiO$_2$ mainly occurs in the visible light and UV region with an energy larger than 2.5 eV, which is attributed to the intrinsic transition from the O$-2p$ orbital to Ti$-3d$ orbital. Upon formation of an interface, the optical absorption edges extend  to the low-energy range, displaying the obvious enhancement of photoactivity under the visible and near-infrared light irradiation. In high-energy region ($>$ 2.5 eV), 1L-BP/TiO$_2$(110) also display better light absorption performance than that of TiO$_2$ substrate. Generally, over the entire sunlight spectrum, the light absorption activity has an enhancement in the whole region ($<$ 4eV), and especially the presence of considerable visible response are located at the infrared and near-infrared light region. The similar enhanced photoactivities are  also found in the case of a Al- and Cl-doped interfaces. The absorbance from low to high exhibits the following order: Al@1L-BP/TiO$_2$ $>$ 1L-BP/TiO$_2$ $>$ Cl@1L-BP/TiO$_2$.

It is known that the functional parametrized by the Perdew, Burke, and Ernzerhof (PBE)\cite{perdew_self-interaction_1981} usually underestimates the band gap and is unable to describe the reduced charge screening and the enhanced electron-electron correlation. The GW and HSE06\cite{heyd_hybrid_2003, heyd_erratum} calculations are used to correct PBE band gaps of phosphorene and TiO$_2$(110). As shown in Table 1, for phosphorene, the G$_0$W$_0$ band gap is closer to the experimental value (2.2 eV).  As for the Al- and Cl-doped cases, their enlarged quasi-particle band gaps are 2.44 and 2.47 eV, respectively.  While, both of G$_0$W$_0$ and PBE results fail to obtain the accurate CBM and VBM of TiO$_2$(110);\cite{migani_quasiparticle_2014} nonetheless, the HSE06 xc-functional reproduces the VBM and CBM levels to within 0.3 eV comparable to the experimental values.\cite{migani_quasiparticle_2014} Therefore, the HSE06 CBM of TiO$_2$(110) is used to match with G$_0$W$_0$ CBM of phosphorene layers.
Assuming that phosphorene layers is the main absorber in the XSCs, the exciton binding energy ($E_b$) in the phosphorene layers is a key quantity which determines the energetics of photoexcited electron transfer to TiO$_2$(110). To address this point, the optical CBM  of the phosphorene is calculated using GW plus Bethe-Salpeter equation (BSE) approach,\cite{deslippe_berkeleygw:_2012, rohlfing_electron-hole_2000} and thus accounting for $E_b$, as show in Figure 3a. This combined scheme utilizing the optical CBM of the donor and the HSE06 VBM of the acceptor takes into account the minimum energy of the exciton formed after photoabsorption in the  donor materials, as well as the electronic quasiparticle level for the transfer of a photoexcited electron to the acceptor. The $\Delta$$\emph{E}_C$ values derived at this combined level of theory (BSE $\Delta$$\emph{E}_C$) are shown in Figure 3b, which yields the same qualitative trends as the PBE results concerning the type-II band alignment.

The light absorbance based on GW plus random phase approximation (GW+RPA) (without electron-hole interaction but including self-energy effects at the GW level) and  plus BSE (GW+BSE) (without electron-hole  interaction) is shown in Figure 4. We focus in this comparison  in the range of 1.0$-$3 eV, which is of key relevance to photovoltaics.  For pure phosphorene, the first prominent peak corresponds to a bright excition with a  $E_b$  of 0.75 eV, locating at 1.28 eV (defined as the optical band gap), in close agreement with the experimental result (1.30 eV),\cite{wang_highly_2015} further highlighting the accuracy of our approach. The absorbance of Al- and Cl-doped phosphorene is plotted in Figure 4c$-$f. The first prominent peaks are  located at 1.62 and 1.62 eV for Al@1L-BP and Cl@1L-BP, corresponding to two bright excitions with $E_b$ of 0.82 and 0.85 eV, respectively, having a slightly increase of 0.07 eV and 0.10 eV compared to pure 1L-BP. To understand the origin of the variation of $E_b$, we calculated the dielectric function with $\omega$ = 0 ($\epsilon_0$). The $\epsilon_0$ exhibits the following order: Cl@1L-BP $<$ Al@1L-BP $<$ pure phosphorene, indicating the Coulomb screening effects in Al$-$P bond and Cl$-$P bond in Al@1L-BP are higher than  that in P$-$P bond of phosphorene. What's more, it¡¯s worth noting that the $E_b$ in pure and doped phosphorene layers show a linear scaling as a function of the quasiparticle (QP) gap (equivalent to G$_0$W$_0$ gap) (Figure 4g), in accordance with previous finding in pure or chemical fictionalized or strained 2D materials,\cite{choi_linear_2015} further extending the applicability of the scaling relationship into doped 2D materials.

As shown in Figure 4c, Al@1L-BP has an optical gap of 1.64 eV, 0.38 eV larger than that of pure phosphorene, and almost the same to Cl@1L-BP (1.62 eV). An absorbance upper limit of $A(\omega)$ $\approx$ 19.5 \% can be reached for these doping cases with only a  small portion of drops compared to perfect phosphorene along the armchair direction. Although doped phosphorene layers require much more energies to achieve the optical excitation  due to the  enlarged optical gaps, their optical CBMs are even higher than the CBM of TiO$_2$(110) (Figure 3a), leading to the larger optical BSE $\Delta$$\emph{E}_C$ of 1.11 and 1.02 eV for Al- and Cl-doped phosphorene, respectively, and thus holding a larger built-in potential to drive exciton separation. These above features can also be quantified by the absorbed photon flux $J_{abs}$, which can be accessed based on the following equation:

\begin{equation}\label{eq:1}
J_{abs} = e\int_{E_g^{opt}}^{\infty}A(\omega)J_{ph}(E)
\end{equation}

where $\mathrm{A}$ is the light absorbance (see in Supporting Information), $J_{ph}(E)$ is the incident photon flux (units of photons/cm$^2\cdot$s$\cdot${eV}), and $\mathrm{E}$ is the photon energy. The  $J_{abs}$ of three interfaces are listed in Table 2, which is expressed as the equivalent short-circuit electrical current density (units of mA/cm$^2$) in the ideal system when every photon is converted to a carrier extracted in a solar cell, so that $J_{abs}$  represents the upper limit for the contribution of the donor material to the short-circuit current in a solar cell.\cite{bernardi_extraordinary_2013} The $J_{abs}$ for the incident light polarized along the x (armchair) direction is larger than along zigzag direction by 1.5$-$2.6 mA/cm$^2$ in phosphorene layers, indicating the anisotropy of light absorption.  The $J_{abs}$ of Al@1L-BP along two directions are larger than that in pure phosphorene, while the $J_{abs}$ of Cl@1L-BP has the oppositive effect, showing hole doping can enlarge the $J_{abs}$, which helps to optimize the light absorption efficiency of phosphorene via doping. The exciting  phenomenon that Al@1L-BP with an enhanced light absorbance by 30 \% compared to pure phosphorene  can be attributed to the following two factors: (i) a noteworthy absorbance peak along the armchair direction is located at 2.25 eV, 0.6 eV higher than the first peak in pure phosphorene, strengthening  Al@1L-BP's armchair absorbance in the energy range of more than 2.0 eV (Figure 4a and c); (ii) for the absorbance in zigzag direction, the light absorption edges has a big red-shift by 1.1 eV from $\sim$2.8 eV to 1.68 eV (Figure 4d), almost close to the one with incident light polarized along the armchair direction, which is very beneficial to further improve the absorption activities in the key energy range for photovoltaics. The tunable $J_{abs}$ in the range of 1.1$-$5.8 mA/cm$^2$, up to 1 order of magnitude larger than nanometer-thick Si, GaAs, and P3HT (in the range of 0.1$-$0.3 mA/cm$^2$) (Table 2). Thus, like perfect phosphorene, doping monolayers also show high efficiency of light absorption.

The photo-induced electron injection, relaxation, and electron$-$hole recombination  at the interface of the hybrid system affect the charge carrier lifetime, and in turn, solar cell current and performance. Previous studies mainly focus on dye sensitized solar cell (DSSC), in which photoexcited electron transfer to TiO$_2$ plays a crucial role in the performance and efficiency of the DSSC.\cite{duncan_ab_2005,duncan_time-domain_2007,craig_trajectory_2005,li_electron_2010} Using trajectory surface hopping methods\cite{tully_molecular_1990,hammesschiffer_proton_1994,parandekar_mixed_2005} implemented within the time-dependent Kohn$-$Sham theory,\cite{fischer_regarding_2011,craig_trajectory_2005} the simulations of photo-induced electron transfer (ET) dynamics at three interfaces were calculated and presented in Figure 5a-c.
For intact phosphorene, about 52$\%$ of the photoexcited states are localized on TiO$_2$ after photo-excitation. The electron transfer process is dominated by the adiabatic electron transfer with  a small portion of the non-adiabatic process. By the exponential fitting, the photoinduced electron transfer from phosphorene  into the TiO$_2$ surface occurs on a 6.1 fs time scale. After Al or Cl doping in phosphorene, we can see the obvious difference in electron dynamics. For the initial PE states, Al doping can increase the initial coupling with a larger percentage ($\sim$60$\%$) of photoexcited states localized on TiO$_2$ (Figure 5b), while Cl doping decreases the ratio of photo-induced electron states on TiO$_2$ to 41$\%$ (Figure 5c). The fitted total electron injection time ($\tau_{inj}$) for intact 1L-BP/TiO$_2$ is 6.1 fs, indicating the ultrafast electron injection process. Al@1L-BP has only a slight influence on the electron transfer with a $\tau_{inj}$ of 7.7 fs, comparable to that of the intact phosphorene, while Cl@1L-BP slows down the $\tau_{inj}$ to 11.0 fs. Both intact and doped interfaces can realize ultrafast electron-hole separation in such short time scale of several femtosecond, on the same order of magnitude as conventional DSSC\cite{duncan_ab_2005,duncan_time-domain_2007,craig_trajectory_2005,li_electron_2010}, and much longer than the injecting time (about 160 fs) in vdW's graphene/TiO$_2$ interface.\cite{ni_review_2007,long_photo-induced_2012} Such ultrafast ET efficiency would be significantly beneficial for the practical application in XSC.

The electron$-$hole recombination at the phosphorene/TiO$_2$ interface occurs by a nonradiative transition of the photoexcited electron (PE) from the CB of TiO$_2$ surface to VBM of phosphorene layers. The electron-hole recombination dynamics of injected electrons are shown in Figure 5d.  At the beginning, almost all electrons are localized on TiO$_2$ substrate, and then transfer back to the VBM of phosphorene layers. For the intact interface, the electron-hole recombination can take place in a recombination time ($\tau_{rec}$) of 0.58 ps obtained from the linear approximation, which is much slower than that of PE injection. Doping Cl and Al can slow down $\tau_{rec}$ to 0.69 ps and 1.52 ps. The large $\tau_{rec}$ can reduce the energy losses and increase the performance of photovoltaic cells, which is very beneficial for maintaining the photocurrent and finally achieving the high power conversion efficiency (PCE) in phosphorene/TiO$_2$ heterostructures. Meanwhile, we can see that the proper doping in phosphorene layers can effectively tune the interfacial electron-hole dynamics and finally the performance of the device. Based on above electron dynamics results,  we can approximately estimate the internal quantum efficiency (IQE, namely the fraction of absorbed photons extracted as carriers at the contacts) by using  $\tau_{rec}$ as the upper limit of carrier lifetimes in phosphorene layers based on the ratio of $\tau_{inj}$ and $\tau_{rec}$, IQE = $1-\tau_{inj}$/$\tau_{rec}$.  The calculated IQE, 1L-BP/TiO$_2$(110),  Al@1L-BP/TiO$_2$(110), and  Cl@1L-BP/TiO$_2$(110) XSCs are 98.9$\%$, 98.9$\%$, and 99.3$\%$, respectively, showing a ultrahigh charge separation efficiency for each of three XSCs.  The estimated IQEs imply that almost the whole absorbed photons contribute to the current, resulting in the short-circuit current $J_{\mathrm{SC}}$ $\approx$ $J_{\mathrm{abs}}$ in the solar cell device.

Although the thermodynamic efficiency limit for thermal carriers  in the absence of nonradiative recombination is set based on the optical gap of the donor through the Schockley$-$Quisser limit,\cite{gregg_photoconversion_2005} the practical PCE for the three heterostructures are more useful than ultimate thermodynamic limits when they come to practical implementation in XSCs. We compute the PCE under AM1.5G illumination by dividing the product $J_{\mathrm{SC}}\times V_{\mathrm{OC}}\times FF$ through the incident power of 100 mW/cm$^2$.\cite{lunt_practical_2011}. Using an mediate open circuit voltage $V_{\mathrm{OC}}$ = 0.6 V, FF values of 0.65, the PCE values of 1.67, 1.71 and 1.22 \% are achieved in 1L-BP/TiO$_2$(110), Al@1L-BP/TiO$_2$(110) and Cl@1L-BP/TiO$_2$(110), respectively, indicating the Al@1L-BP/TiO$_2$(110) has the highest PCE among the three interfaces. Combined with the obtained PCE values as derived above, we estimate three heterostructures with a thickness of 1.0 nm would achieve a power density of 16.7, 17.1 and 12.2 MW/L, respectively,  higher by approximately 1$-$3 orders of magnitude compared to existing solar cells, such as 1 $\mu$m thick GaAs with a power density of 290 kW/L, and 35 $\mu$m thick Si of 5.9 kW/L.\cite{bernardi_extraordinary_2013} These PCE values are comparable to MoS$_2$/graphene and MoS$_2$/WS$_2$ bilayer ultrathin photovoltaic (PV) with the PCE of $\sim$0.1$-$1.5$\%$.\cite{bernardi_extraordinary_2013} Although these PCEs  at atomistic level are much lower than conventional GaAs or Si PV devices with PCEs more than 20 $\%$, thicker multilayer stacking (50-100 nm thick) in a bulk heterojunction manner\cite{halls_efficient_1995, yu_polymer_1995} could be carried out to maximize the interface area and PCEs, which may improve the PCE by a factor of dozens of times compared to that in GaAs (see in Table 2).

In summary, we have systemically investigated the electronic and optical properties of heterostructures consisting of phosphorene and doped monolayers interfaced with TiO$_2$(110) surface. These heterojunctions show type-II band alignments, enhanced photoactivities and  large built-in potential for carrier separation. Doping in phorsphorene has a well tunability on the excitonic binding energy, optical band gap, light absorbance, electron-hole dynamics as well as power conversion efficiency in a excitonic solar cell (XSC). The idea of XSC based on phosphorene$-$TiO$_2$ heterosturcutres are proposed, where the phosphorene layer is served as the donor  and the TiO$_2$(110) as the acceptor. Three heterostructures (1L-BP/TiO$_2$(110), Al@1L-BP/TiO$_2$(110) and Cl@1L-BP/TiO$_2$(110)) XSCs show tunable power conversion efficiency in the range of 1.22$-$1.71$\%$ and ultrahigh power densities, dosens of times higher than conventional  GaAs solar cells, comparable with MoS$_2$/WS$_2$ XSC. The nonadiabatic molecular dynamics within the time-dependent density functional theory framework shows ultrafast electron transfer of 6.1$-$10.8 fs, and slow electron$-$hole recombination of 0.58$-$1.08 ps, further ensuring the practical power conversion efficiencies in XSC.  These features, including the tunable optical band gap, type-II interface band alignment, high optical absorbance, large power conversion efficiency enable phosphorene-based heterostructures being promising for next-generation flexible optoelectronic devices. Our results presented here may stimulate further efforts on the rational design of future solar cell devices based on the combinations of 2D materials and 3D wide band gap semiconductors.

\begin{suppinfo}
Methods, simulation Details, and the PDOS of Al@1L-BP/TiO$_2$(110) and Cl@1L-BP/TiO$_2$(110).
\end{suppinfo}

\section{Notes}
The authors declare no competing financial interests.


\begin{acknowledgement}
L.Z and B.S acknowledges financial support from BremenTRAC$-$COFUND fellowships, co-financed by the Marie Curie Program of the European Union. A.D, L.Z and T.F acknowledge financial support from Australian Technology Network (ATN) and Deutscher Akademischer Austausch Dienst (DAAD) German Academic Exchange Service. This work is partially financial supported from MOST (grant 2012CB921403) and NSFC (grant 11222431). We thank Gang Lu at California State University Northridge for help  with electronic dynamics calculations. The support of the Supercomputer Center of Northern Germany (HLRN Grant No. hbp00027) is  also acknowledged.
\end{acknowledgement}

\clearpage
\newpage

\begin{figure}
\centering
\includegraphics [width=13cm]{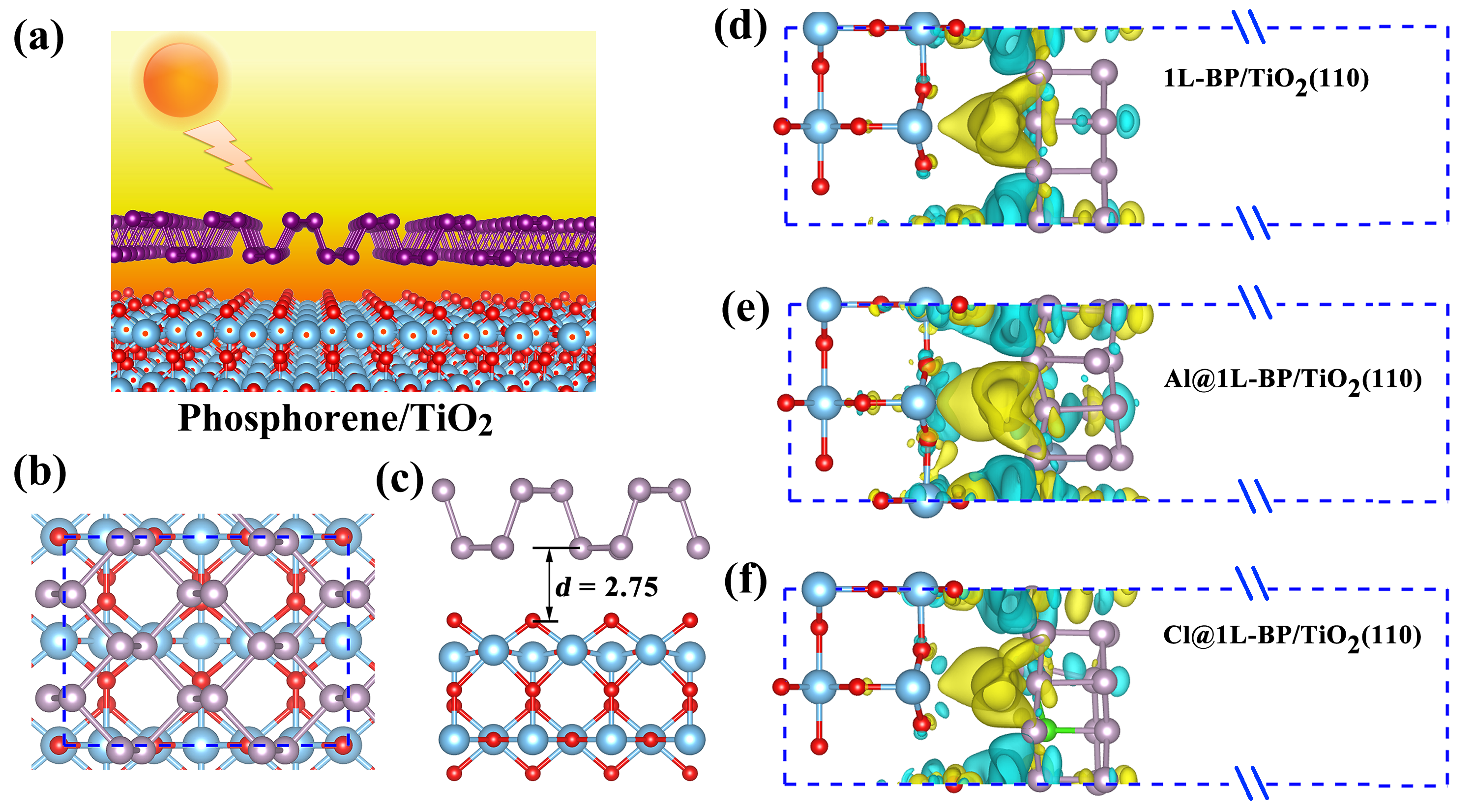}\\
\caption{(a) Schematic drawing of the interfaces between phosphorene layer and TiO$_2$(110).  The optimized interfaces between 1L-BP and TiO$_2$(110) (denoted as 1L-BP/TiO$_2$(110)): (a) top view; (b)side view. Gray: P; red: O; Skyblue: Ti. The distance between 1L-BP and TiO$_2$(110) in (b) is inserted. panels (d)-(f): Charge density differences for a 1L-BP/TiO$_2$(110), Al@1L-BP/TiO$_2$(110), and Cl@1L-BP/TiO$_2$(110), respectively. The yellow region represents charge accumulation, and the cyan region indicates charge depletion; the isosurface value is 0.0004 e/{\AA}$^3$. }
\label{Fig. 1}
\end{figure}

\begin{figure}
\centering
\includegraphics [width=13cm]{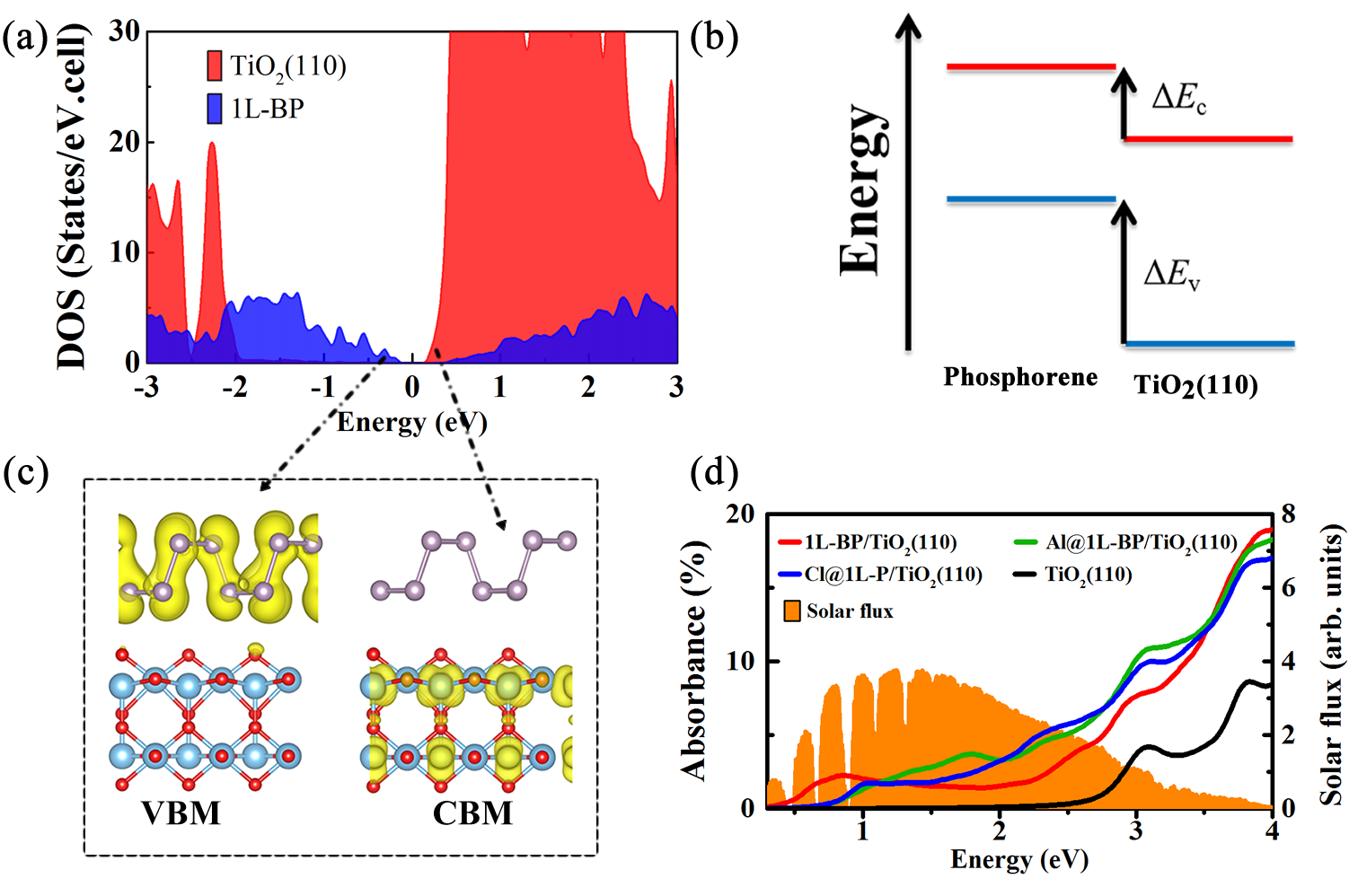}\\
\caption{(a) Density of states (DOS) for 1L-BP/TiO$_2$(110) interface. The Fermi level is set to zero. (b) Band alignment at a 1L-BP/TiO$_2$(110) interface, as predicted using DFT. Valence ($\Delta$$\emph{E}_V$) and conduction ($\Delta$$\emph{E}_C$ ) band offsets at 1L-BP/TiO$_2$(110) interface are referenced, respectively, to the valence and conduction band edges of the acceptor. (c) VBM and CBM charge density contours (in 0.0014 e/{\AA}) of 1L-BP/TiO$_2$(110) interface, respectively. (d) Absorbance of three heterostructures, 1L-BP/TiO$_2$(110), Al@1L-BP/TiO$_2$(110), Cl@1L-BP/TiO$_2$(110), as well as TiO$_2$(110) surface, overlapped to the incident AM1.5G solar flux.}
\label{Fig. 2}
\end{figure}

\begin{figure}
\centering
\includegraphics [width=8cm]{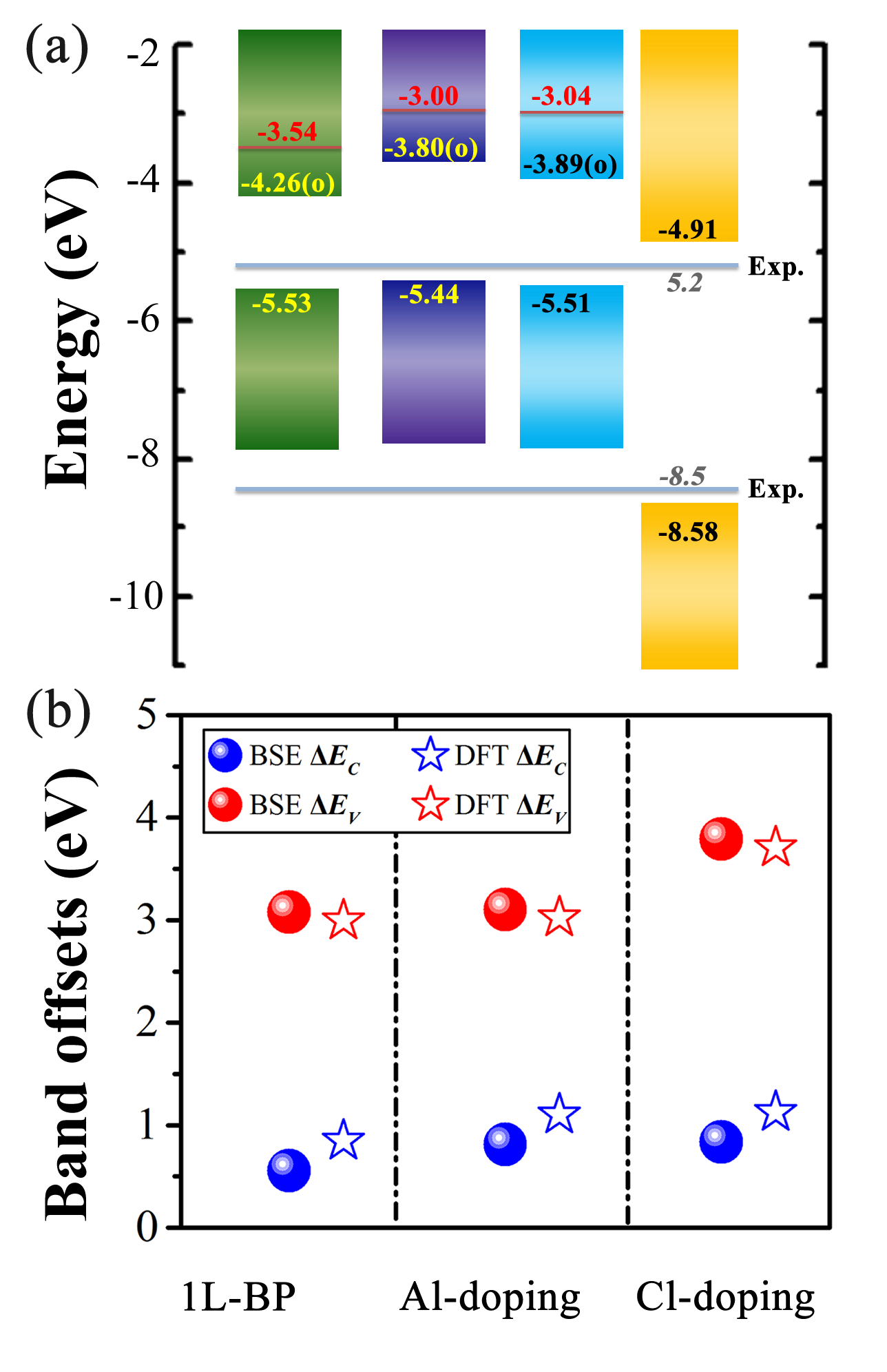}\\
\caption{(a) Variation of valence band maximum (VBM) and conduction band minimum (CBM) with respect to the vacuum level for pristine,  and Al-, Cl-doped phosphorene from G$_0$W$_0$ and G$_0$W$_0$ plus BSE (GW+BSE) calculations, as well as the TiO$_2$ structures from HSE06 using PBE xc-functionals level.  The letter ``o"  in parentheses means optical CBM of phosphorene layers obtained from GW+BSE calculations. Gray regions denote VBM and CBM energies derived from
the experimental results. (b) Valence band offsets ($\Delta$$\emph{E}_V$) and conduction band offsets ($\Delta$$\emph{E}_C$ ) based on  GW+BSE and DFT calculations at phosphorene-TiO$_2$(110) interfaces for perfect and doping monolayers, respectively.  $\Delta$$\emph{E}_V$ and $\Delta$$\emph{E}_C$ are referenced to the valence band maximum (VBM) and conduction band minimum (CBM) of the acceptor (TiO$_2$(110)), respectively. The positive $\Delta$$\emph{E}_V$ and $\Delta$$\emph{E}_C$ indicate type-II band alignment.}
\label{Fig. 3}
\end{figure}

\begin{figure}
\centering
\includegraphics [width=13cm]{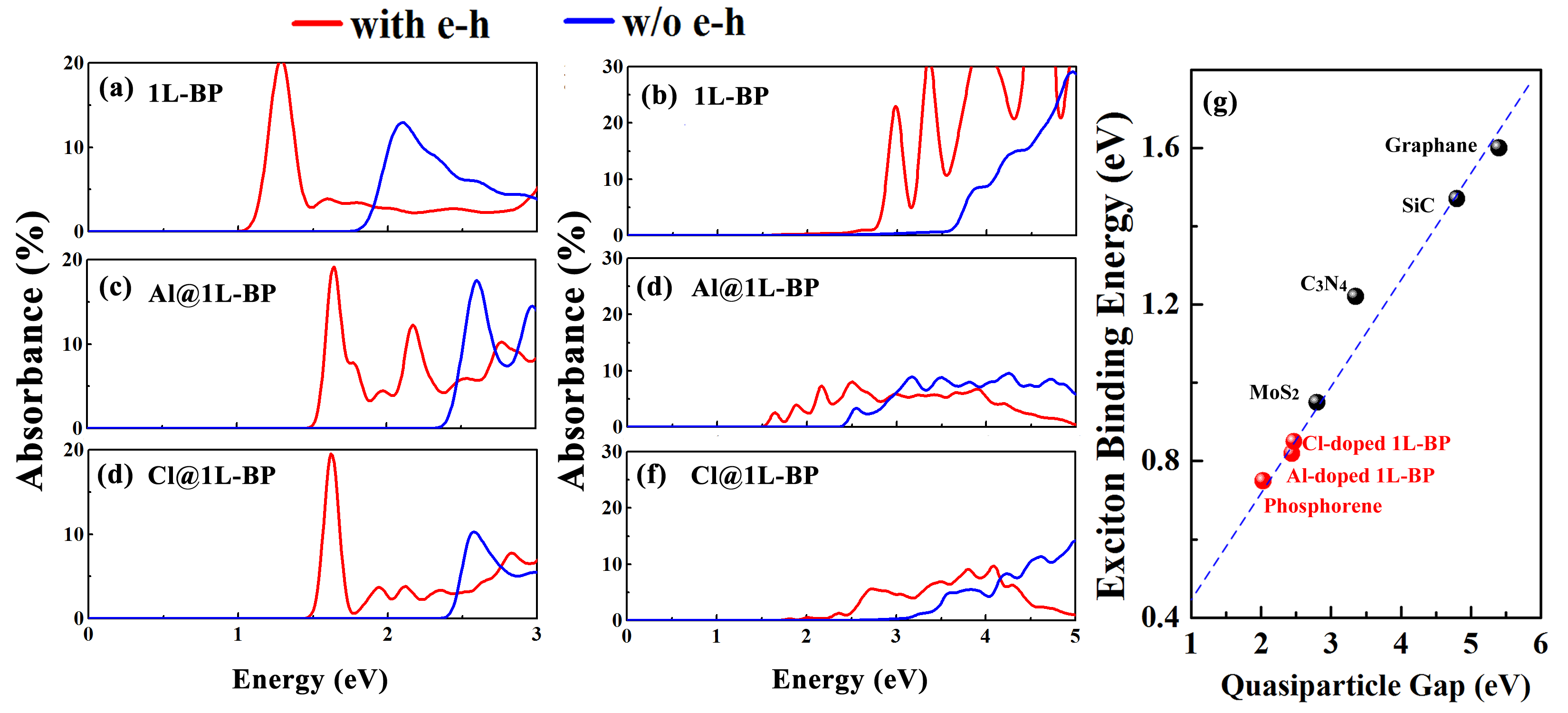}\\
\caption{Absorbance spectra of (a),(b) pristine, and (c),(d) Al-doped, and (e),(f) Cl-doped phosphorene for the incident light polarized along the x (armchair) direction and along the y (zigzag) direction. The single-particle optical absorption without \emph{e}$-$\emph{h} interaction are presented by blue solid lines while those spectra with e-h interaction included are presented by red solid lines. The spectra are broadened by Lorentzians with line widths of 0.05 eV. Interband electron transition from the VB to the CB of phosphorene; after, this photogenerated electron is able to transfer to the bottom of the CB of TiO$_2$(110), which need to overcome the binding energy of exciton generated in donor materials. (g) The exciton binding energy ($E_b$) versus the QP band gap ($E_g$) for various representative 2D materials. The dashed line represents the fitted linear relation in the form of $E_b={\alpha}E_g +\beta$, with $\alpha$ =0.23 and $\beta$ = 0.45. $E_b$ of SiC,\cite{choi_linear_2015} C$_3$N$_4$,\cite{wei_many-body_2013} Graphane,\cite{wei_strong_2013} MoS$_2$\cite{qiu_optical_2013} monolayers are obtained from relevant  references.}
\label{Fig. 4}
\end{figure}

\begin{figure}
\centering
\includegraphics [width=13cm]{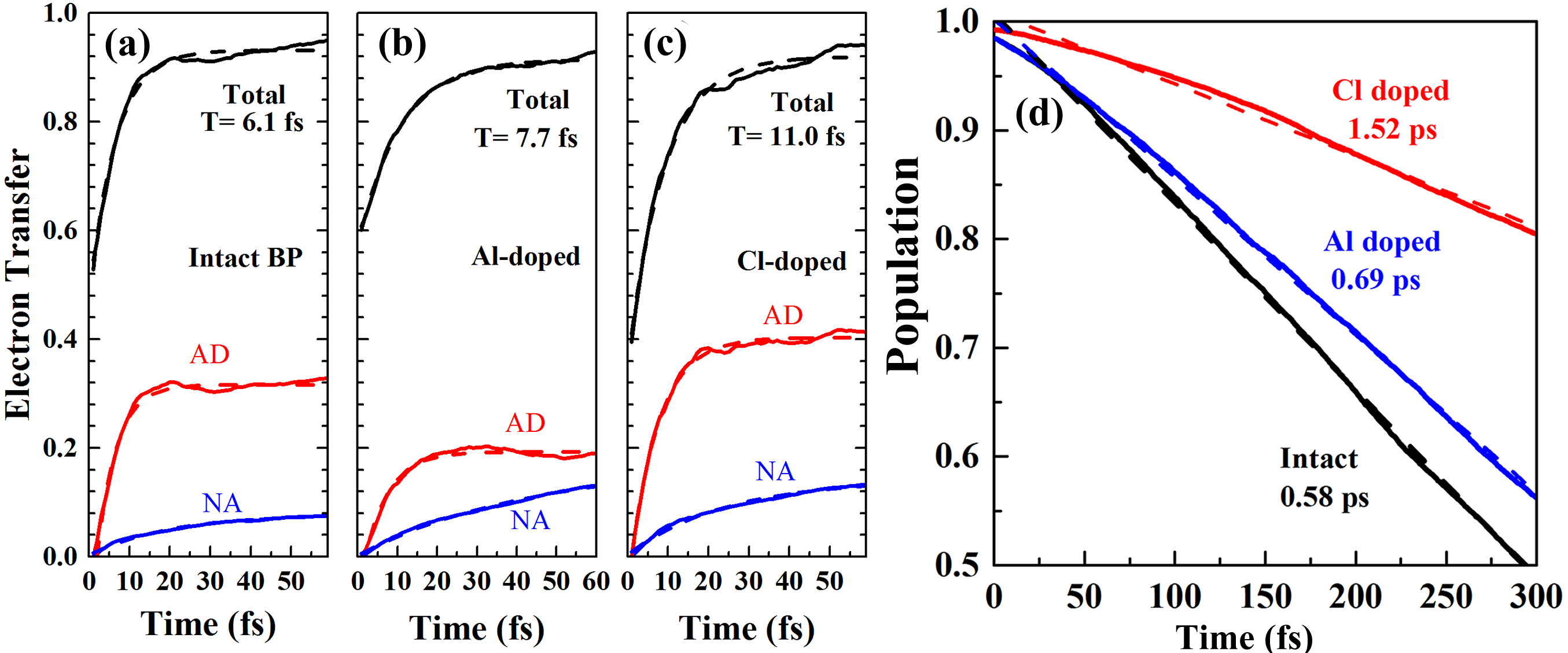}\\
\caption{Average electron transfer dynamics of photoexcited states for (a) intact, (b) Cl doped,  (c) Al doped phosphorene. The solid black, solid red and solid blue lines represent the total adiabatic and nonadaiabatic electron transfer.  Thee total, adiabatic (AD) and nonadiabatic (NA) ET are indicated by the black solid, red solid and blue solid lines, respectively. The dashed lines show the exponential fits of each lines. (d) Electron-hole recombination dynamics from phosphorene CBs to TiO$_2$ highest VBs in three heterostructures. Intact, Al- and Cl-doped interfaces are  indicated in black solid, blue solid, red solid line, respectively. The dashed lines are fitted by linear functions. The oped phosphorene can slightly slow down the process of injected electron migrated back to TiO$_2$.}
\label{Fig. 5}
\end{figure}

\clearpage
\newpage
\begin{table}
\caption{\label{tab:1} Band Offsets (in eV), Exciton Binding Energies ($E_b$), the First Optical Absorption Peak (''Optical gap'', $E_{g}^{o}$), G$_0$W$_0$ Energy Gaps of the Phosphorene and their Al- and Cl-doped Monolayers, as well as IQE and PCE When Interfaced with TiO$_2$(110).}
\begin{center}
\begin{tabular}{l c c c c }
\hline
 Compound           & 1L-BP     & Al@1L-BP     & Cl@1L-BP   & TiO$_2$(110)\\
\hline
   CBM              &$-$3.54      & $-$3.00      & $-$3.04     &$-$4.91$^a$ \\
   optical CBM      &$-$4.26      &$-$3.8        & $-$3.89     &$-$4.91$^a$ \\
   built-in potential  &0.65    & 1.11          & 1.02         &na \\
   VBM              &$-$5.53	    &$-$5.44	 &$-$5.51      &$-$8.58$^a$\\
   $E_b$       & 0.72      & 0.82           &0.85       & na  \\
   PBE gap          & 0.84      & 1.01          & 1.18      & 2.17 \\
   Optical gap   & 1.28      & 1.62          & 1.62      & na  \\
   G$_0$W$_0$ Gap   & 2.03      & 2.44          &2.47       & 4.26  \\
\hline
$^a$ obtained from HSE06 calculation &  &  &  & \\
\end{tabular}
\end{center}
\end{table}

\begin{table}
\caption{\label{tab:2} Optical Band Gaps of Donor Material, and Absorbed Photon Flux $J_{abs}^a$ (incident light polarized along the armchair direction), Flux $J_{abs}^z$ (along zigzag direction), total $J_{abs}$  under AM1.5G Solar Illumination, PCE, and power density (PD) for the 1L-BP/TiO$_2$(110), Al@1L-BP/TiO$_2$(110) and Cl@1L-BP/TiO$_2$(110). Computed Using Equation 1 with the Absorbance Values in Figure 4a$-$f $^a$ }
\begin{minipage}{15cm}
\begin{tabular}{l c c c c c c c}
\hline
 Material      &$E_{g}^{o}$ (eV) & $J_{abs}^{a}$ &$J_{abs}^{z}$  & $J_{abs}$ (mA/cm$^2$) & thickness &PCE & PD(kW/L) \\
\hline
   Phosphorene &1.28         & 4.3                      &2.0               & 3.15 & 1nm &   1.67\% &  16 700   \\
   Al-doped    &1.38         & 4.41                     &2.53              & 3.47 &1nm  & 1.71\%&     17 100  \\
   Cl-doped    &1.38         & 3.15                    &1.01               &  2.08& 1nm  &  1.22\%&    12 200  \\
   Si          &1.11         &                         &                   &0.1   & 35 $\mu$m &   $\sim$29\%&  5.9   \\
   GaAs        &1.42         &                          &                  &0.3  & 1 $\mu$m &   20.6\% &      290   \\
\hline
\footnotetext[1]{$J_{abs}$ quantifies the flux of absorbed photons, converted to units of equivalent electrical current. The same quantities are also shown for 1 nm thick representative bulk materials in ultrathin PV, taken from the literature.\cite{bernardi_extraordinary_2013}}
\end{tabular}
\end{minipage}
\end{table}
\clearpage
\newpage
\bibliography{bp-tio2}

\begin{tocentry}

\includegraphics [width=9cm] {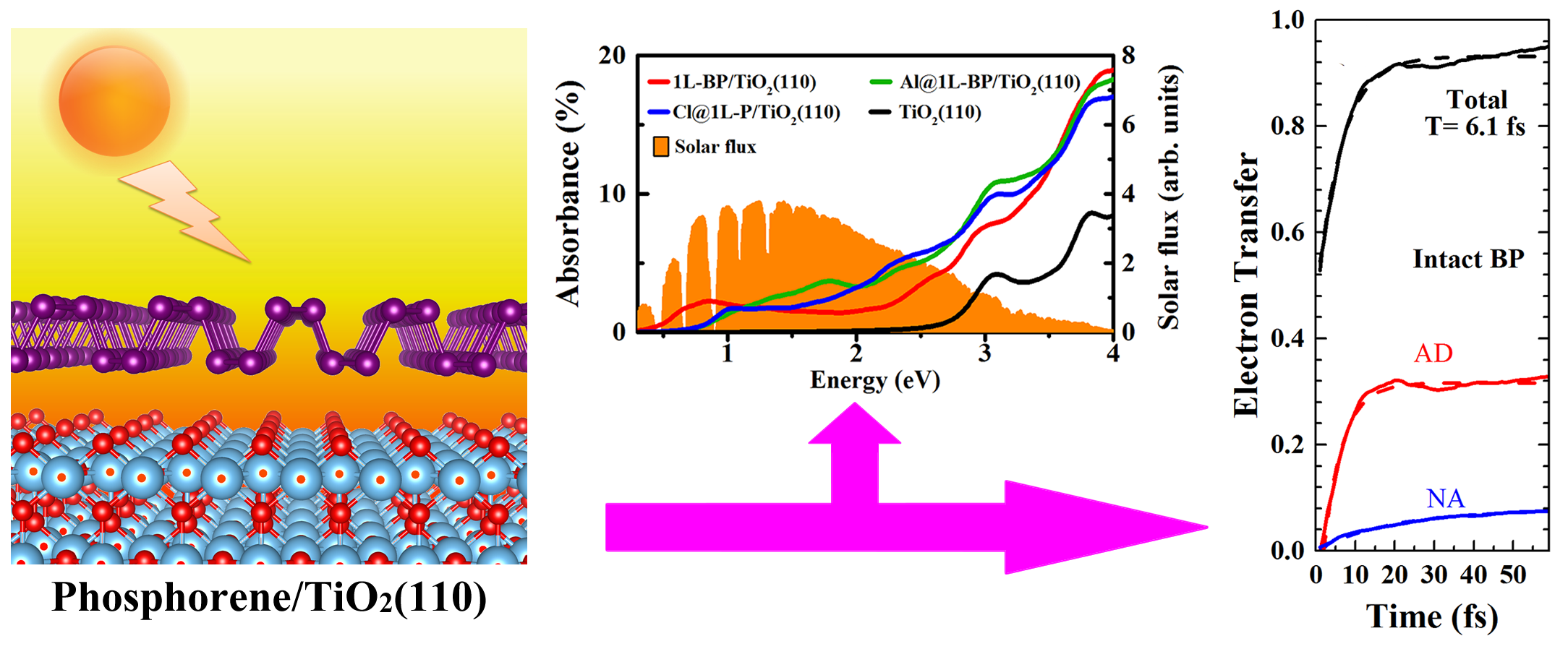}

     Novel excitonic solar cells (XSCs) based on pure phosphorene and doped monolayers interfaced with TiO$_2$ are proposed. These heterostructures show type-II band alignment and enhanced light absorbance. Doping in phosphorene has a tunability on built-in potential, charge transfer, light absorbance, which helps to optimize the light absorption efficiency of phosphorene. These heterostructures used as active layers in a XSC can attain high power conversion efficiencies and ultrahigh power densities, comparable with MoS$_2$/WS$_2$ XSC  at atomistic thickness, and dozens of times higher than convectional solar cells. Ultrafast electron transfer of 6.1$-$10.8 fs, and slow electron$-$hole recombination of 0.58$-$1.08 ps, further guaranteeing the practical power conversion efficiencies in XSC
\end{tocentry}

\end{document}